\RequirePackage{fix-cm}
\documentclass[smallextended]{svjour3}       
\smartqed  
\usepackage{mathrsfs}
\usepackage{amsfonts}
\usepackage{amstext}
\usepackage{amsmath}
\usepackage{amssymb}
\usepackage{bm}
\usepackage{CJK}
\usepackage{bbm}
\usepackage[dvips]{graphicx}

\usepackage{graphicx}
\usepackage{graphics}

\usepackage{amsmath}
\usepackage{color}

\definecolor{darkred}  {rgb}{0.5,0,0}
\definecolor{darkblue} {rgb}{0,0,0.5}
\definecolor{darkgreen}{rgb}{0,0.5,0}
\usepackage{hyperref}
\hypersetup{
    pdftitle = {QRT Proposal},
    pdfauthor = {},
    colorlinks = true,
    urlcolor  = blue,         
    linkcolor = red,     
    citecolor = blue,    
    filecolor = darkred       
}
\usepackage{mathtools}
\def\ra{\rangle}
\def\la{\langle}
\def\bb{\mathbb}
\def\ot{\otimes}

\newcommand{\bea}{\begin{eqnarray}}
\newcommand{\eea}{\end{eqnarray}}
\newcommand{\be}{\begin{equation}}
\newcommand{\ee}{\end{equation}}
\newcommand{\ba}{\begin{equation}\begin{aligned}}
\newcommand{\ea}{\end{aligned}\end{equation}}

\newcommand{\beax}{\begin{eqnarray*}}
    \newcommand{\eeax}{\end{eqnarray*}}
\newcommand{\bex}{\begin{equation*}}
\newcommand{\eex}{\end{equation*}}

\def\be{\begin{equation}}
\def\ee{\end{equation}}

\newcommand{\mH}{\mathcal{H}}

\newcommand{\mP}{\mathcal{P}}

\newcommand{\mS}{\mathcal{S}}

\newcommand{\tr}{{\rm Tr}}

\newcommand{\mE}{\mathcal{E}}
\newcommand{\mC}{\mathcal{C}}

\def\>{\rangle}
\def\<{\langle}

\begin{document}

\title{Complete $k$-partite entanglement measure
}

\titlerunning{Complete $k$-partite entanglement measure}        

\author{Jinxing Zhao \and Yu Guo \and Fei He
}

\authorrunning{J. Zhao et al.} 

\institute{\at
              School of Mathematical Sciences, Inner Mongolia University, Hohhot, Inner Mongolia 010021, People's Republic of China}

\date{Received: date / Accepted: date}

\maketitle

\begin{abstract}
The $k$-partite entanglement, which focus on at most how many particles in the global system are entangled but separable from other particles, is complementary to the $k$-entanglement that reflects how many splitted subsystems are entangled under partitions of the systems in characterizing multipartite entanglement. Very recently, the theory of the complete $k$-entanglement measure has been established in [Phys. Rev. A 110, 012405 (2024)]. Here we investigate whether we can define the complete measure of the $k$-partite entanglement. Consequently, with the same spirit as that of the complete $k$-entanglement measure, we present the axiomatic postulates that a complete $k$-partite entanglement measure should require. Furthermore, we present two classes of $k$-partite entanglement measures and show that one is complete while the other one is unified but not complete except for the case of $k=2$. 

\end{abstract}


\section{Introduction}

In 2005, G\"{u}hne \textit{et al}. introduced the $k$-partite entanglement in Ref.~\cite{Guhne2005njp}. It is closely related to the $k$-producible state: if a quantum state is not $k$-producible, it is termed $(k+1)$-partite entangled. While the $k$-entanglement reflects how many splitted subsystems are entangled under partitions of the systems~\cite{Guo2024pra,Hong2012pra}, the $k$-partite entanglement concentrate on at most how many particles in the global system are entangled but separable from other particles~\cite{Guhne2005njp,Hong2023epjp,Li2025aqt}. It has been shown that $k$-producibility plays a crucial role in both quantum nonlocality~\cite{Curchod2015pra,Liang2015prl,Lin2019pra,Lu2018prx} and quantum metrology~\cite{Toth2014jpa}. Particularly, $k$-producibly entangled states for larger $k$ exhibit higher sensitivity in phase estimation~\cite{Hyllus2012pra,Gessner2018prl,Qin2019npj}.

Clearly, these two ways of exhibiting entanglement, i.e.,  the $k$-entanglement and the $k$-partite entanglement, are complementary to each other in characterizing the multipartite entanglement which remains challenging to understand undeniably since the complexity increases substantially with the number of parties~\cite{Horodecki2009,Verstraete2003pra,Gour2010prl,Jungnitsch2011prl,Xie2021prl,Jacob2021prl,Li2024pra,Guo2020pra,Guo2022jpa,Guo2022entropy}. Recently, the $k$-partite entanglement measure based on concurrence have been presnted~\cite{Hong2023epjp,Li2025aqt}. Very recently, we established the theory of the complete $k$-entanglement measure in Ref.~\cite{Guo2024pra}. It was shown that, in the framework of the complete measure of quantum correlation, the distribution of the correlation could be depicted exhaustively since the correlation could be compared not only between the global system and the subsystem (or the systems under arbitrary partition) but also between different subsystems (or the systems under arbitrary partition)~\cite{Guo2024pra,Guo2020pra,Guo2022entropy,Guo2024rip,G2021qst,Guo2023pra}.
Along this line, the aim of this paper is to discuss how can we establish the axiomatic postulates for the complete $k$-partite entanglement measure.

The rest of the paper is arranged as follows. We review the concept of $k$-partite entanglement, the $k$-partite entanglement measures proposed in Ref.~\cite{Hong2023epjp,Li2025aqt}, and the coarsening relation of multipartite partitions in Sec.~\ref{sec2}. In Sec.~\ref{sec3}, we present the definition of the complete $k$-partite entanglement measure, and then give two general ways of constructing $k$-partite entanglement measures and discuss whether they are complete in Sec.~\ref{sec4}. Sec. \ref{sec5} lists some examples of $k$-partite entanglement measures according to the methods in Sec.~\ref{sec4}. Finally, in Sec. \ref{sec6}, we summarize the results of the paper.

\section{Notations and Preliminaries}\label{sec2}

For convenience of discussing the complete measure of the $k$-partite entanglement in the next sections, we review some basic notations and terminologies in Sec~\ref{2.1}, and introduce the $k$-partite entanglement measures proposed in literature so far in Sec.~\ref{2.2}. We then introduce the coarsening relation of the multipartite partitions which is necessary when we discuss the completeness of a multipartite quantum correlation measure (also see in Ref.~\cite{Guo2022entropy,Guo2024rip}).

We fix some notations first. We denote by $A_1A_2\cdots A_n$ an $n$-partite quantum system. Let $X_1|X_2|\cdots |X_m$ be an $m$-partition of $A_1A_2\cdots A_n$ (for instance, partition $AB|C|DE$ is a $3$-partition of the 5-particle system $ABCDE$ with $X_1=AB$, $X_2=C$ and $X_3=DE$. The case of $m=n$ is just the original $n$-particle system without any other partition, namely, $A_1A_2\cdots A_n$ means $A_1|A_2|\cdots |A_n$. So $m< n$ in general unless otherwise specified). We denote by $\varDelta(X_t)$ the number of subsystems contained in $X_t$, for instance, for the 3-partition $AB|C|DE$ of $ABCDE$, $\varDelta(X_1)=\varDelta(AB)=2$, $\varDelta(X_2)=\varDelta(C)=1$ and $\varDelta(X_3)=\varDelta(DE)=2$. If $\varDelta(X_t)\leqslant  k$ for any $1\leqslant  t\leqslant  m$, we call it a $k$-fineness partition. We denote by $\Gamma_{\!k}^f$ the set of all $k$-fineness partitions of the given system $A_1A_2\cdots A_n$.

\subsection{$k$-partite entanglement}\label{2.1}

\begin{figure}
	\centering
	\includegraphics[width=84mm]{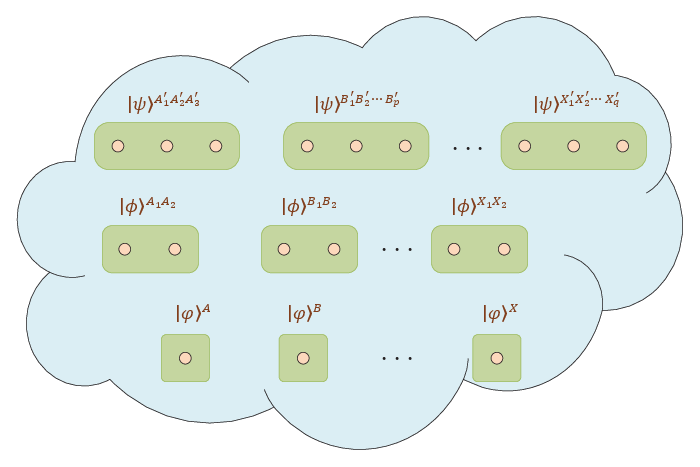}\\
	{(a)}
	\vspace{3mm}\\
	\includegraphics[width=60mm]{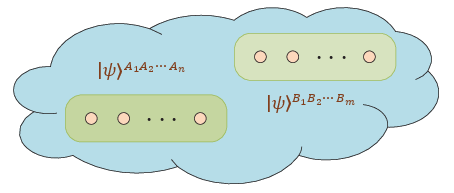}	\\
	{(b)}
	\vspace{2mm}
	\caption{\label{fig1}(color online). (a) $3$-partite entangled pure state $|\Psi\ra=|\psi\ra^{A'_1A'_2A'_3}|\psi\ra^{B'_1B'_2\cdots B'_p} \cdots|\psi\ra^{X'_1X'_2\cdots X'_q}|\phi\ra^{A_1A_2}|\phi\ra^{B_1B_2}\cdots|\phi\ra^{X_1X_2}|\varphi\ra^A|\varphi\ra^B\cdots$ $|\varphi\ra^X$, where $|\psi\ra^{A'_1A'_2A'_3}$, $|\psi\ra^{B'_1B'_2\cdots B'_p}$, $\dots$, $|\psi\ra^{X'_1X'_2\cdots X'_q}$ are genuinely entangled states, $3\leqslant p\leqslant q$, $|\phi\ra^{A_1A_2}$, $|\phi\ra^{B_1B_2}$, $\dots$, $|\phi\ra^{X_1X_2}$ are entangled states. In fact, if one of $|\psi\ra^{A'_1A'_2A'_3}$, $|\psi\ra^{B'_1B'_2\cdots B'_p}$, $\dots$, $|\psi\ra^{X'_1X'_2\cdots X'_q}$ is genuinely entangled, $|\Psi\ra$ is also $3$-partite entangled. Here we just take the general form of a $3$-partite entangled pure state.
		(b) $|\Phi\ra=|\psi\ra^{A_1A_2\cdots A_k}|\psi\ra^{B_1B_2\cdots B_l}$ with $k$, $l\geqslant 0$, $k+l\geqslant3$, is a $3$-entangled pure state if one of the following is true: (i) $|\psi\ra^{A_1A_2\cdots A_k}$ and $|\psi\ra^{B_1B_2\cdots B_l}$ are genuinely entangled states, $k$, $l\geqslant3$, (ii)  $|\psi\ra^{A_1A_2\cdots A_k}$ and $|\psi\ra^{B_1B_2\cdots B_l}$ are entangled states, $k=l=2$, (iii) If $k=0$ or $l=0$, $|\Phi\ra$ is genuinely entangled.}
\end{figure}

A pure state $|\psi\ra$ of an $n$-partite system $A_1A_2\cdots A_n$ with state space $\mH^{A_1A_2\cdots A_n}$ is called $k$-producible ($1\leqslant  k\leqslant  n-1$), if it can be represented as~\cite{Guhne2005njp}
\bea\label{k-producible}
|\psi\rangle=|\psi\rangle^{X_1}|\psi\ra^{X_2}\cdots|\psi\ra^{X_m}
\eea
under some $k$-fineness partition $X_1|X_2|\cdots |X_m$ of $A_1A_2\cdots A_n$. Let $\mS^X$ be the set of all density operators acting on the state space $\mH^X$. For mixed state $\rho\in\mS^{A_1A_2\cdots A_n}$, if it can be written as a convex combination of $k$-producible pure states, i.e., $\rho=\sum\limits_ip_i|\psi_i\rangle\langle\psi_i|$ with $|\psi_i\ra$s are $k$-producible, it is called $k$-producible, where the pure state $|\psi_i\rangle$s might be $k$-producible in different $k$-fineness partitions. If a quantum state is not $k$-producible, it is termed $(k+1)$-partite entangled. Note that if $|\psi\ra$ admits the form as in Eq.~\eqref{k-producible}, it is called $m$-separable. $|\psi\ra$ is $m$-entangled if it is not $m$-separable. An $n$-partite mixed state $\rho$ is $m$-separable if it can be written as $\rho=\sum_{i}p_i|\psi_i\rangle \langle\psi_i|$ with $|\psi_i\ra$s are $m$-separable, wherein the contained $\{|\psi_i\rangle\}$ can be $m$-separable with respect to different $m$-partitions. Otherwise, it is called $m$-entangled. By definition, the $k$-partite entanglement is different from the $k$-entanglement in general, but they are equivalent only in some special cases. For example, the $n$-partite entangled state is just the genuine multipartite entangled state and the one-producible state coincides with the fully separable state. If $|\psi\ra^{ABC}$ is a genuine entangled state, then $|\psi\ra^{ABC}|\psi\ra^{D}|\psi\ra^{E}|\psi\ra^F$ is four-separable and three-partite entangled state. Also note that, a state of which some reduced state of $m$ parties is genuinely entangled, contains $m$-partite entanglement, but not vice versa in general~\cite{Guhne2005njp}. For more clarity, we compare $3$-partite entangled pure state with $3$-entangled pure state in Fig.~\ref{fig1}.

A pure state $|\phi\rangle$ is said to be genuinely $k$-producible (or genuinely $k$-partite entangled) \cite{Guhne2005njp} if it is $k$-producible but not $(k-\!1)$-producible. A mixed state $\rho\in\mS^{A_1A_2\cdots A_n}$ is genuinely $k$-producible if it is $k$-producible and for any $k$-producible pure states ensemble of $\rho$, $\rho=\sum_ip_i|\psi_i\ra\la\psi_i|$, there is at least one $|\psi_i\rangle$ is genuinely $k$-producible.

Let $\mS_{P(k)}$ $(k=1,2,\ldots,n-1)$ denote the set of $k$-producible quantum states in $\mS^{A_1A_2\cdots A_n}$ and $\mS_{P(n)}:=\mS$. It follows that
\bea 
\mS_{P(1)} \subset \mS_{P(2)}\subset\cdots\subset \mS_{P(n-1)}\subset \mS_{P(n)},
\eea
$\mS\setminus\!\mS_{P(k)}$ is the set consisting of all $(k+1)$-partite entangled states, and $\mS_{P(k)}\!\!\setminus\!\mS_{P(k-1)}$ is the set of all genuinely $k$-producible states.

\subsection{$k$-partite entanglement measures via $q$-concurrence and $\alpha$-concurrence} \label{2.2}

A positive function $E_{(k)}:\mS^{A_1A_2\cdots A_n}\longrightarrow\bb{R}_+$ is called a $k$-partite entanglement measure ($k$-PEM) if it fulfills: 
(i) $E_{(k)}(\rho)=0$ for any $\rho\in \mS_{P(k-1)}$ and $E_{(k)}(\rho)>0$ for any $\rho\in \mS\setminus \mS_{P(k-1)}$,
(ii) $E_{(k)}(\rho)$ does not increase under $n$-partite local operations and classical communication (LOCC), namely, $E_{(k)}(\varepsilon(\rho))\leq  E_{(k)}(\rho)$ for any $n$-partite LOCC $\varepsilon$. Item (ii) guarantees that $E_{(k)}$ is invariant under local unitary operation.
In addition, a $k$-PEM $E_{(k)}$ on $\mS^{A_1A_2\cdots A_n}$ is convex and non-increasing on average under 
$n$-partite LOCC, it is call a $k$-partite entanglement monotone ($k$-PEMo). A $k$-PEM/$k$-PEMo $E_{(k)}$ is called a genuine $k$-PEM/$k$-PEMo if $E_{(k)}(\rho)>0$  but $E_{(k+1)}(\rho)=0$ for any genuinely $k$-partite entangled state $\rho$.

Hong \textit{et al.} presented a $k$-PEMo in Ref.~\cite{Hong2023epjp} via concurrence. For any pure state $|\psi\ra\in\mH^{A_1A_2\cdots A_n}$, the $k$-PEMo was defined as~\cite{Hong2023epjp}
\bea\label{k-part-entanglement-C0}
C_{(k)}(|\psi\rangle)=\min\limits_{\Gamma_{\!k-\!1}^f}\frac{\sum_{t=1}^{m}\sqrt{2[1-{\rm Tr}(\rho_{X_t}^{2})]}}{m},
\eea
where $\rho_{X_t}={\rm Tr}_{{\overline X}_t}(|\psi\rangle\langle\psi|)$, ${\overline X}_t$ is the complement of subsystem $X_t$, the minimum is taken over all the $(k\!-\!1)$-fineness partitions in $\Gamma_{\!k-\!1}^f$.
For mixed states, it is defined by the convex-roof extension, i.e.,
\beax
C_{(k)}(\rho)=\min\limits_{p_i, |\psi_i\ra}\sum_ip_iC_{(k)}(|\psi_i\ra),
\eeax
where the minimum runs over all ensembles $\{p_i, |\psi_i\ra|\rho=\sum_ip_i|\psi_i\ra\la\psi_i|\}$. In what follows, we give only the measures for pure states, for the case of mixed states they are all defined by the convex-roof extension with no further statement. Obviously, any measure that is defined in this way is convex. 

Very recently, Li \textit{et al.} proposed two $k$-PEMos in Ref.~\cite{Li2025aqt}. For any pure state $|\psi\ra\in\mH^{A_1A_2\cdots A_n}$, the $k$-PEMo via the $q$-concurrence was defined as~\cite{Li2025aqt}
\bea\label{k-part-entanglement-C1}
C_{q(k)}(|\psi\rangle)=\min\limits_{\Gamma_{\!k-\!1}^f}\sqrt{\frac{\sum_{t=1}^{m}[1-{\rm Tr}(\rho_{X_t}^{q})]}{m}},
\eea
and the $k$-PEMo via the $\alpha$-concurrence was expressed by~\cite{Li2025aqt}
\bea\label{k-part-entanglement-C2}
C_{\alpha(k)}(|\psi\rangle)=\min\limits_{\Gamma_{\!k-\!1}^f}\sqrt{\frac{\sum_{t=1}^{m}[{\rm Tr}(\rho_{X_t}^{\alpha})-1]}{m}},
\eea
where the minimum is taken over all the $(k\!-\!1)$-fineness partitions in $\Gamma_{\!k-\!1}^f$. Note here that, the notations here are different from $E_{q-k}$ and $E_{\alpha-k}$ in Ref.~\cite{Li2025aqt} ($C_{q(k+1)}=E_{q-k}$, $C_{\alpha(k+1)}=E_{\alpha-k}$). Ref.~\cite{Li2025aqt} also gave the following two $k$-PEMos:
\bea\label{k-part-entanglement-GC1}
C_{G,q(k)}(|\phi\rangle)=\left( \frac{\prod\limits_{\gamma_i\in\Gamma_{\!k-\!1}^f}\left[ \sum\limits_{t=1}^{m_i}(1-{\rm Tr}\rho_{X_{t(i)}}^q)\right] }{\prod_{i=1}^{\left| \Gamma_{\!k-\!1}^f\right| }m_{i}}\right) ^{\frac{1}{2\left| \Gamma_{\!k-\!1}^f\right| }}
\eea
and 
\bea\label{k-part-entanglement-GC2}
C_{G,\alpha(k)}(|\phi\rangle)=\left( \frac{\prod\limits_{\gamma_i\in\Gamma_{\!k-\!1}^f}\left[ \sum\limits_{t=1}^{m_i}({\rm Tr}\rho_{X_{t(i)}}^{\alpha}-1)\right] }{\prod_{i=1}^{\left| \Gamma_{\!k-\!1}^f\right| }m_{i}}\right) ^{\frac{1}{2\left| \Gamma_{\!k-\!1}^f\right| }},~~~
\eea
where $\rho_{X_{t(i)}}$ is the reduced density operator with respect to subsystem ${X_{t(i)}}$, and $m_i$ refers to $\gamma_i$ is a $m_i$-partition, $\left| \Gamma_{\!k-\!1}^f\right|$ is the cardinal number of $\Gamma_{\!k-\!1}^f$. The notations in Eqs.~\eqref{k-part-entanglement-GC1},~\eqref{k-part-entanglement-GC2} are different from $\varepsilon_{q-k}$ and $\varepsilon_{\alpha-k}$ in Ref.~\cite{Li2025aqt} ($C_{G,q(k+1)}=\varepsilon_{q-k}$, $C_{G,\alpha(k+1)}=\varepsilon_{\alpha-k}$).

\subsection{Coarsening relation of multipartite partitions}

Let $X_1|X_2| \cdots |X_{k}$ and $Y_1|Y_2| \cdots |Y_{l}$ be two partitions of $A_1A_2\cdots A_n$ or subsystem of $A_1A_2\cdots A_n$, $k\leq  n$, $l\leq  n$. We denote by~\cite{Guo2024pra,Guo2022entropy}
\bea
X_1|X_2| \cdots| X_{k}\succ^a Y_1|Y_2| \cdots |Y_{l},\\
X_1|X_2| \cdots| X_{k}\succ^b Y_1|Y_2| \cdots |Y_{l},\\
X_1|X_2| \cdots| X_{k}\succ^c Y_1|Y_2| \cdots |Y_{l}~
\eea 
if $Y_1|Y_2| \cdots |Y_{l}$ can be obtained from $X_1|X_2| \cdots| X_{k}$ by 
\begin{itemize}
	\item[(a)] Discarding some subsystem(s) of $X_1|X_2| \cdots| X_{k}$,
	\item[(b)] Combining some subsystems of $X_1|X_2| \cdots| X_{k}$,
	\item[(c)] Discarding some subsystem(s) of some subsystem(s) $X_t$ provided that $X_{t}=A_{t(1)}A_{t(2)}\cdots A_{t(f(t))}$ with $f(t)\geqslant2$, $1\leqslant t\leqslant k$,
\end{itemize}
respectively. For example,
\beax 
&&A|B|C|D\succ^a A|B|D\succ^a B|D,\\
&&A|B|C|D\succ^b AC|B|D\succ^b AC|BD, \\
&&A|BC\succ^c A|B.
\eeax
Here, we denote $ABCD$, $ABD$, $BD$ and $AB$ by $A|B|C|D$, $A|B|D$, $B|D$ and $A|B$, respectively. 

For any subsystem of $A_1A_2\cdots A_n$ with arbitrary partition, it can always be derived from the global system via the coarsening relations (a)-(c) or some of them. So, based on these three coarsening relations, we can analyze not only the information relation between any subsystem and the global subsystem but also the information relation between any subsystems in detail. For instance, based on these coarsening relations, we have established the complete global entanglement measure~\cite{Guo2020pra,Guo2024rip}, the complete genuine entanglement measure~\cite{Guo2022entropy,Guo2024rip}, the complete multipartite quantum discord~\cite{G2021qst}, the complete multipartite quantum mutual information~\cite{Guo2023pra} and the complete $k$-entanglement measure~\cite{Guo2024pra}. Furthermore, we discussed the complete monogamy relation of these measures~\cite{Guo2024pra,Guo2020pra,Guo2022entropy,Guo2024rip,G2021qst,Guo2023pra},
where exploring the monogamy relation of the quantum correlations is one of the fundamental tasks in the quantum resource theory~\cite{Guo2020pra,G2021qst,Coffman,Pawlowski,streltsov2012are,Augusiak2014pra,Osborne,Dhar,GG,GG2019}.

\section{Completeness of the $k$-PEM} \label{sec3}

When we deal with the various quantum correlations living in a multipartite system, the most quintessential relation in a multipartite system is indeed the the coarsening relation, i.e., the coarsening relations of type (a)-(c). 
The ``completeness'' of a measure for multipartite quantum correlation mainly refers to that there is a unified criterion for quantifying different subsystems or systems under different partition, which means the amount of the quantum correlations contained in different particles or particles under arbitrary partition can be compared with each other consistently and compatibly~\cite{Guo2024pra,Guo2020pra,Guo2022entropy,G2021qst,Guo2023pra}. It makes up for the previous bipartite measure which can only quantify the quantum correlation under the given bipartite splitting. By reviewing the key point in defining a complete measure of quantum correlation~\cite{Guo2024pra,Guo2020pra,Guo2022entropy,G2021qst,Guo2023pra}, we can conclude that there are two steps to reveal such a completeness of a given measure: the first step is the unification condition which is mainly related to the coarsening relation of type (a), and the second one is the hierarchy condition which is determined by the coarsening relation of type (b).

We now give the axiomatic postulates for the unified $k$-PEM and the complete $k$-PEM based on the coarsening relation of the partitions of the system. Hereafter, $E_{(k)}(X)$ denotes $E_{(k)}(\rho^X)$.
A $k$-PEM $E_{(k)}$ is called \textit{unified} if it satisfies the unification condition: (i) (symmetry) $E_{(k)}({A_1A_2\cdots A_n})=E_{(k)}({A_{\pi(1)}A_{\pi(2)}\cdots A_{\pi(n)}})$ for all $\rho\in\mS^{A_1A_2\cdots A_n}$ and any permutation $\pi$ of $\{1, 2, \cdots, n\}$; (ii) (additivity) $E_{(k)}(A_1A_2\cdots A_r\ot A_{r+1}A_{r+2}\cdots A_n)=E_{(k)}(A_1A_2\cdots A_r)+E_{(k)}(A_{r+1}A_{r+2}\cdots A_n)$ holds for all $\rho^{A_1A_2\cdots A_r}\ot\rho^{A_{r+1}A_{r+2}\cdots A_n}$; (iii) ($k$-monotone)
\bea\label{k-monotone}
E_{(k)}(A_1A_2\cdots A_n)\leqslant E_{(k-1)}(A_1A_2\cdots A_n)
\eea
holds for all $\rho\in\mS^{A_1A_2\cdots A_n}$, $k\geqslant3$; and (iv) (coarsening monotone)
\bea\label{k-coarsen}
E_{(k)}(X_1|X_2| \cdots| X_{p})\geqslant E_{(k)}(Y_1|Y_2| \cdots |Y_{q})
\eea
holds for all states $\rho\in\mS^{A_1A_2\cdots A_n}$ whenever $X_1|X_2| \cdots| X_{p}\succ^a Y_1|Y_2| \cdots |Y_{q}$ with $k\leqslant q\leqslant p$. Item (i) is clear, i.e., the symmetry is an inherent feature of entanglement measure indeed. The $k$-partite entanglement contained in $A_1A_2\cdots A_r\ot A_{r+1}A_{r+2}\cdots A_n$ is composed of two parts, i.e., $A_1A_2\cdots A_r$ and $A_{r+1}A_{r+2}\cdots A_n$. So we demand item (ii). If a state is $(k-1)$-producible, it must be $k$-producible, but not vice versa, so we require condition (iii). For the generalized $n$-qudit GHZ state $\frac{1}{\sqrt{d}}(|00\cdots 0\ra+|11\cdots1\ra+\cdots|d-1\ra|d-1\ra\cdots|d-1\ra)$, Eq.~\eqref{k-coarsen} is always true for any $k$-PEM. In addition, $E_{(k)}(|\psi\ra^{A_1A_2\cdots A_k}|\psi\ra^{A_{k+1}}\cdots|\psi\ra^{A_n})\geqslant E_{(k)}(\rho^{A_1A_2\cdots A_{k-1}}\ot|\psi\ra\la\psi|^{A_{k+1}}\cdots|\psi\ra\la\psi|^{A_n})=0$ for any $|\psi\ra^{A_1A_2\cdots A_k}|\psi\ra^{A_{k+1}}\cdots|\psi\ra^{A_n}$, $\rho^{A_1A_2\cdots A_{k-1}}=\tr_{A_k}|\psi\ra\la\psi|^{A_1A_2\cdots A_k}$. Therefore item (iv) is straightforward from this point of view. Hereafter, if a $k$-PEM $E_{(k)}$ obeys Eq.~\eqref{k-monotone} and Eq.~\eqref{k-coarsen}, we call it is \textit{$k$-monotonic} and \textit{coarsening monotonic}, respectively.

A unified $k$-PEM $E_{(k)}$ is called \textit{complete} if it satisfies the hierarchy condition additionally: (v) (tight coarsening monotone) 
\bea\label{w-k-t-coarsen}
E_{(k)}(A_1A_2\cdots A_n)\geqslant E_{(k)}(Y_1|Y_2| \cdots |Y_{q})
\eea
holds for all state $\rho\in\mS^{A_1A_2\cdots A_n}$ whenever $A_1A_2\cdots A_n\succ^b Y_1|Y_2|\cdots |Y_{q}$ such that, for any $i$, either $\rho^{Y_i}$ is pure or $\rho^{Y_i}$ is the reduced state of some genuinely entangled pure state (or entangled bipartite pure state), $1\leqslant i\leqslant q<n$. If a $k$-PEM $E_{(k)}$ satisfies Eq.~\eqref{w-k-t-coarsen}, we call it is \textit{tightly coarsening monotonic}. For example, take $|\psi\ra=|\psi\ra^{ABCD}|\psi\ra^{EF}|\psi\ra^{GH}|\psi\ra^{I}$ with $|\psi\ra^{ABCD}$ is genuinely entangled, $|\psi\ra^{EF}$ and $|\psi\ra^{GH}$ are entangled, then the partition $AB|CD|EF|GHI$ is such a case since $|\psi\ra^{EF}$ and $|\psi\ra^{GHI}$ are pure states, $\rho^{AB}$ and $\rho^{CD}$ are reduced states of the genuine pure state $|\psi\ra^{ABCD}$. That is, at least intuitively, such a coarsening operation can not generate $k$-partite entanglement. So we hope Eq.~\eqref{w-k-t-coarsen} should be satisfied. In addition, we take
\beax 
|\psi\ra=|\psi\ra^{ABC}|\psi\ra^{DE}|\psi\ra^{F}|\psi\ra^{GH}|\psi\ra^{IJ},
\eeax 
then 
\beax 
E_{(3)}(|\psi\ra)=E_{(3)}(|\psi\ra^{ABC}).
\eeax
But
\beax 
E_{(3)}(A|B|C|D|EFGI|H|J)
=E_{(3)}(ABC)+E_{(3)}(D|EFGI|H|J)
\eeax 
which is larger than $E_{(3)}(|\psi\ra)$ whenever $|\psi\ra^{DE}$, $|\psi\ra^{GH}$ and $|\psi\ra^{IJ}$ are entangled states. Namely, Eq.~\eqref{w-k-t-coarsen} is violated under the general coarsening operation of type (b). We thus adjust the hierarchy condition (i.e., the tight coarsening monotone condition) as Eq.~\eqref{w-k-t-coarsen} in stead of the ``strict'' hierarchy condition in Refs.~\cite{Guo2024pra,Guo2020pra}.

One need note here that, for any given $k$-PEM $E_{(k)}$,   
\bea\label{hierachy3}
E_{(k)}(X_1|X_2|\cdots|X_{p})\geqslant E_{(k)}(X'_1|X'_2| \cdots |X'_{p})
\eea 
holds for any $\rho\in\mS^{A_1A_2\cdots A_n}$ whenever $X_1|X_2| \cdots| X_{p}\succ^c X'_1|X'_2| \cdots |X'_{p}$ since $\rho^{X'_1|X'_2| \cdots |X'_{p}}$ is obtained from $\rho^{X_1|X_2| \cdots| X_{p}}$ by a partial trace and such a partial trace is indeed a $p$-partite LOCC, $1\leqslant k\leqslant p< n$.

We take the 4-partite system $ABCD$ for example. $E_{(4)}$ is $k$-monotonic means
\beax 
E_{(4)}(ABCD)\leqslant E_{(3)}(ABCD)\leqslant E_{(2)}(ABCD)
\eeax
for any $\rho\in\mS^{ABCD}$, and $E_{(3)}$ is coarsening monotonic refers to
\beax
&&E_{(3)}(ABCD)\geqslant E_{(3)}(ABC),\\
&&E_{(3)}(ABCD)\geqslant E_{(3)}(ABD),\\
&&E_{(3)}(ABCD)\geqslant E_{(3)}(ACD),\\
&&E_{(3)}(ABCD)\geqslant E_{(3)}(BCD),
\eeax
for any state $\rho\in\mS^{ABCD}$. If $E_{(3)}$ is tightly coarsening monotonic,
\beax
&&E_{(3)}(ABCD)\geqslant E_{(3)}(A|B|CD),\\
&&E_{(3)}(ABCD)\geqslant E_{(3)}(A|BC|D),\\
&&E_{(3)}(ABCD)\geqslant E_{(3)}(AB|C|D),\\
&&E_{(3)}(ABCD)\geqslant E_{(3)}(AC|B|D),\\
&&E_{(3)}(ABCD)\geqslant E_{(3)}(AD|B|C),\\
&&E_{(3)}(ABCD)\geqslant E_{(3)}(A|C|BD)
\eeax
for any genuinely entangled pure state $|\psi\ra\in\mH^{ABCD}$.

\section{Two classes of $k$-PEMos}\label{sec4}

In this section, we give two classes of $k$-PEMos, where the first class is based on the unified multipartite entanglement measure introduced in Refs.~\cite{Guo2020pra,Guo2022jpa,Guo2024rip} and the second class is similar to that of the $k$-entanglement measure defined by the minimal sum of the reduced functions in Ref.~\cite{Guo2024pra}. Hereafter, if a measure $E$ of entanglement for bipartite pure state is defined via some function of the reduced state, i.e., $E(|\psi\ra^{AB})=h(\rho^A)$, such a function $h$ is called the reduced function of $E$~\cite{Guo2024pra}. For example, the concurrence of $|\psi\ra^{AB}$ is defined by $C(|\psi\ra^{AB})=\sqrt{2(1-\tr\rho_A^2)}$, then $h_C(\rho)=\sqrt{2(1-\tr\rho^2)}$ is the reduced function of $C$.

\subsection{$k$-PEMo from unified multipartite entanglement measure}

If $|\psi\ra=|\psi\ra^{AB}|\psi\ra^{CDE}|\psi\ra^{FGH}|\psi\ra^I$ with $|\psi\ra^{CDE}$ and $|\psi\ra^{FGH}$ are genuinely entangled, then the 3-partite entanglement is only contained in $|\psi\ra^{CDE}$ and $|\psi\ra^{FGH}$. The 3-partite entanglement of $|\psi\ra$ can be quantified as $E^{(3)}(|\psi\ra^{CDE})+E^{(3)}(|\psi\ra^{FGH})$ for some unified multipartite entanglement measure $E^{(k)}$ (the unified multipartite entanglement measure (MEM) was introduced in Ref.~\cite{Guo2020pra,Guo2024rip}, e.g., $E^{(n)}(|\psi\ra^{A_1A_2\cdots A_n})=\frac12\sum_iS(\rho^{A_i})$ is a unified MEM, where $S(\rho)=-\tr(\rho\log_2\rho)$ is the von Neumann entropy of $\rho$). Similarly, the 2-partite entanglement should be $E^{(2)}(\psi\ra^{AB})+E^{(3)}(|\psi\ra^{CDE})+E^{(3)}(|\psi\ra^{FGH})$.

In general, for any given pure state $|\psi\ra=|\psi\ra^{A_1A_2\cdots A_n}$ in $\mH^{A_1A_2\cdots A_n}$, we assume it is not $(k-1)$-producible. Then there exists a $l$-fitness partition $X_1|X_2|\cdots|X_m$, $l\geqslant k$, such that 
\bea\label{condition of X_t}
\left\lbrace \begin{array}{l}\!\!\Delta(X_t):=s(t)\geqslant k, \\
	\!\!\rho^{X_t}~\textrm{is a genuinely entangled pure state}
\end{array}\right. ~~~~
\eea
for some subsystem $X_t$ in the partition $X_1|X_2|\cdots|X_m$. Let $t_1$, $t_2$, $\dots$, $t_r$ be all of the subscripts such that $X_{t_i}$ satisfies the condition~\eqref{condition of X_t} corresponding to all possible $l$-fitness partitions with $l\geqslant k$. It turns out that
\bea 
|\psi\ra=|\psi\ra^{X_{t_1}}|\psi\ra^{X_{t_2}}\cdots|\psi\ra^{X_{t_r}}|\phi\ra^{X_*}
\eea
under some permutation of the subsystems, where $X_*$ denotes the subsystem complementary to $X_{t_1}X_{t_2}\cdots X_{t_r}$. In such a sense, we can quantify the $k$-partite entanglement of $|\psi\ra$ by
\bea 
E_{(k)}(|\psi\ra)=\sum\limits_{j=1}^rE^{(s(t_j))}(|\psi\ra^{X_{t_j}})
\eea
for any given unified MEM $E^{(n)}$.

With the notations above, we give the following definition of a $k$-PEMo:
\bea\label{E-k2}
E_{(k)}(|\psi\ra)=\begin{cases}
	\sum\limits_{j=1}^rE^{(s(t_j))}(|\psi\ra^{X_{t_j}}),&s(t_j)\geq k~\text{for some}~j,\\
	0,&\text{otherwise}.
\end{cases}
\eea

\begin{theorem}\label{th2} Let $h$ be a reduced function.
	If $E^{(n)}(|\psi\ra)=\frac12\sum_ih(\rho^{A_i})$, then $E_{(k)}$ is a unified $k$-PEMo, and moreover, $E_{(k)}$ is a complete $k$-PEMo whenever $h$ is subadditve.
\end{theorem}

\begin{proof} 
	Items (i) and (ii) are straightforward. For any $|\psi\ra\in\mS^{A_1A_2\cdots A_n}$, we suppose that $E_{(k)}(|\psi\ra)=\sum\limits_{j=1}^rE^{(s(t_j))}(|\psi\ra^{X_{t_j}})>0$ for some partition $X_{t_1}|X_{t_2}|\cdots|X_{t_r}|X_*$. It turns out that $E_{(k)}(|\psi\ra^{X_{t_j}})=E_{(k-1)}(|\psi\ra^{X_{t_j}})$ for any $j$, $E_{(k)}(|\phi\ra^{X_*})=0$, but it is possible that
	$E_{(k-1)}(|\phi\ra^{X_*})>0$, which implies (iii) is true. For any partition $X_1|X_2|\cdots|X_p$ of $A_1A_2\cdots A_n$, $p\geqslant k$, we consider $X_2|\cdots|X_p$ w.n.l.g., namely the partition that by discarding $X_1$ from $X_1|X_2|\cdots|X_p$. If $X_1$ is $X_{t_j}$ or some subsytem(ssome) of $X_{t_j}$, (iv) is clear since $E^{(s)}$ is unified~\cite{Guo2024rip} [a unified MEM is decreasing under the coarsening relation of type (a)]. If $X_1$ is $X_*$ or some subsytem(s) of $X_*$, (iv) is clear since $E_{(k)}(|\psi\ra)=E_{(k)}(\rho^{X_2|\cdots|X_p})$. If $X_1$ cotains some subsytem(s) of some $X_{t_j}$, (iv) is also ture since $E^{(s)}$ is unified. The other cases can be argued similarly.

	If the reduced function $h$ is subadditive, then $E^{(n)}$ is complete~\cite{Guo2020pra,Guo2024rip}, which means that $E^{(n)}$ is nonincreasing under the coarsening operation of type (b). This completes the proof.
\end{proof}

Another candidate for the unified global MEM is the one defined by the sum of all bipartite entanglement~\cite{Guo2022jpa}, i.e.,
\bea\label{sum2}
 \qquad \mE^{(n)}(|\psi\ra^{A_1A_2\cdots A_n})
=\left\lbrace \begin{array}{ll}
	\frac{1}{2}\sum\limits_{i_1\leq \cdots\leq i_s, s< n/2}h(\rho^{A_{i_1}A_{i_2}\cdots A_{i_s}}),& \mbox{if}~n~\mbox{is odd},\\
	\frac{1}{2}\sum\limits_{i_1\leq \cdots\leq i_s<n, s\leq n/2}h(\rho^{A_{i_1}A_{i_2}\cdots A_{i_s}}),& \mbox{if}~n~\mbox{is even},
\end{array}\right. ~~~~~
\eea
where $h$ is some given reduced function. We denote by $\mE_{(k)}$ the quantity that is defined as in Eq.~\eqref{E-k2} just with $\mE^{(s(t_j))}$ replacing $E^{(s(t_j))}$. Using similar arguments as in the proof of Theorem~\ref{th2}, we can conclude the following theorem.

\begin{theorem}\label{th3}
	$\mE_{(k)}$ is a unified $k$-PEMo, and $\mE_{(k)}$ is complete if $h$ is subadditive.
\end{theorem}

In Eq.~\eqref{E-k2}, $|\psi\ra^{X_{t_j}}$ is genuinely entangled, so $E^{(s(t_j))}$ can be any genuine entanglement measure instead. For example, the genuine entanglement measure from the minimal reduced function, which is defined by~\cite{Guo2024rip}
\beax\label{gmin1}
\varepsilon_{g''}^{(n)}(|\psi\ra^{A_1A_2\cdots A_n})
=\min\limits_{X}h(\rho^X)
\eeax
where $h$ is some given reduced function, $X\subsetneq\{A_1,A_2, \dots, A_n\}$, namely the minimum runs over all possible reduced states. Then the corresponding $k$-PEMo is not unified in general since $\varepsilon_{g''}^{(n)}$ may increase under the coarsening relation of type (a)~\cite{Guo2024rip}.

\subsection{$k$-PEMo from the minimal sum}

Let $|\psi\ra=|\psi\ra^{A_1A_2\cdots A_n}$ be a pure state in $\mH^{A_1A_2\cdots A_n}$ and $h$ be a reduced function.  For any $\gamma_i^f\in\Gamma_k^f$, we write 
\bea\label{P_k}
\mP_k^{\gamma_i^f}(|\psi\ra)\equiv\frac12\sum_{t=1}^{m}h(\rho^{X_{t(i)}}),~1\leqslant k<n,
\eea
where $X_{1(i)}|X_{2(i)}|\cdots|X_{m(i)}$ corresponds to $\gamma_i^f$, $\rho^{X}=\tr_{\overline{X}}|\psi\ra\la\psi|$, and $\overline{X}$ denotes the subsystem complementary to those of $X$. The coefficient ``1/2'' is fixed by the unification condition when the measures defined via $\mP_k^{\gamma_i^f}$ are regarded as unified $k$-PEMs. We define
\bea\label{sum}
E'_{(k)}(|\psi\ra)=\min\limits_{\Gamma_{\!k-\!1}^f}\mP_{\!k-\!1}^{\gamma_i^f}(|\psi\ra),
\eea
where the minimum is taken over all feasible $(k\!\!-\!\!1)$-fineness partitions in $\Gamma_{\!k-\!1}^f$. By definition, for any $\rho\in\mS^{A_1A_2\cdots A_n}$, $E'_{(k)}(\rho)>0$ if and only if $\rho$ is $k$-partite entangled.

\begin{theorem}\label{th1}
	$E'_{(k)}$ is a unified $k$-PEMo and $E'_{(2)}$ is complete if the reduced function is subadditive.
\end{theorem}

\begin{proof}
	By definition, it is straightforward that $E'_{(k)}$ is a $k$-PEMo. We show below it satisfies the unification conditions (i)-(iv). We only need to check items (iii) and (iv) since (i) and (ii) are clear. Since $\Gamma^f_{k\!-\!2}\subseteq\Gamma^f_{k\!-\!1}$, this implies (iii) is true. For any given $|\psi\ra\in\mH^{A_1A_2\cdots A_n}$, we assume w.n.l.g. that 
	\beax 
	E'_{(k)}(|\psi\ra)=\frac12\left[ h(\rho^{X_1})+h(\rho^{X_2})+h(\rho^{X_3})\right] 
	\eeax 
	for some 
	$X_1|X_2|\cdots|X_m$. If 	
	\beax Y_1|Y_2| \cdots |Y_{p}\succ^a Y'_1|Y'_2| \cdots |Y'_{q}
	\eeax 
	with $Y'_1|Y'_2| \cdots |Y'_{q}$ is obtained by discarding subsystem(s) $Y_i$s such that $Y_i$s are contained in $X_4X_5\cdots X_m$, Eq.~\eqref{k-coarsen} is clear. If $Y'_1|Y'_2| \cdots |Y'_{q}$ is obtained by discarding subsystem(s) $Y_i$s such that $Y_i$s are contained in $X_1X_2X_3$, there are three subcases: (a) $Y_i=X_1$ (w.n.l.g.), (b) $Y_i=X_{21}$, $X_2=X_{21}X_{22}$ (w.n.l.g.), and (c) $Y_i=X_1X_{21}$ (w.n.l.g.). For the subcase of (a), it turns out that 
	\beax 
	&&E'_{(k)}(|\psi\ra^{X_1X_2X_3})\\
	&=&\frac12\left[ h(\rho^{X_1})+h(\rho^{X_2})+h(\rho^{X_3})\right]\\
	&> & \frac12\sum_ip_i\left[ h(\rho^{X_{2(i)}})+h(\rho^{X_{3(i)}})\right]\\
	&\geqslant & E'_{(k)}(\rho^{X_2X_3})
	\eeax 
	for any pure state ensemble $\{p_i,|\psi\ra^{X_{2(i)}X_{3(i)}}\}$ of $\rho^{X_2X_3}$ since $h$ is concave. That is, Eq.~\eqref{k-coarsen} holds true still. For the subcase of (b), it follows that
	\beax 
	&&E'_{(k)}(|\psi\ra^{X_1X_2X_3})\\
	&\geqslant & \frac12\sum_ip_i\left[ h(\rho^{X_{1(i)}})+h(\rho^{X_{22(i)}})+h(\rho^{X_{3(i)}})\right]
	\geqslant  E'_{(k)}(\rho^{X_{12}X_2X_3})
	\eeax 
	for any pure state ensemble $\{p_i,|\psi\ra^{X_{1(i)}X_{22(i)}X_{3(i)}}\}$ of $\rho^{X_{1}X_{22}X_3}$, where the first inequality holds since discarding $X_{11}$ is a partial trace which is a special LOCC on $X_1|X_2|X_3$, and $E'_{(k)}$ in such a case is $E^{(3)}$ acting on $\mS^{X_1X_2X_3}$ which leads to decreasing under LOCC.
	The subcase of (c) is clear from the arguments for (a) together with (b). That is, Eq.~\eqref{k-coarsen} is valid for both subcases.

	It is clear that the completeness of $E'_{(2)}$ is reduced to the subadditivity of the reduced function. This completes the proof.
\end{proof}

By definition, if the reduced function is subadditive, it can be easily checked that
\bea 
E'_{(k)}(\rho)\leqslant E_{(k)}(\rho)\leqslant \mE_{(k)}(\rho).
\eea 
If $h$ is subadditive, then the minimal partition is the ones that contained in $\Gamma_{\!k-\!1}^f\!\!\setminus\!\Gamma_{\!k-\!2}^f$. $E'_{(k)}$ is not complete in general if $k\geqslant 3$. For example, if 
\beax 
 E'_{(3)}(|\psi\ra^{ABCD}|\psi\ra^{EF}|\psi\ra^{GHI})
=\frac12\left[h(\rho^{AB})+h(\rho^{CD})+h(\rho^{GH})+h(\rho^I) \right], 
\eeax
it follows that
\beax 
&&E'_{(3)}(A|BC|D|E|F|G|H|I)\\
&=&\frac12\left[h(\rho^{A})+h(\rho^{BCD})+h(\rho^{GH})+h(\rho^I) \right] \\
&\geqslant &E'_{(3)}(ABCDEFGH)
\eeax
whenever $h(\rho^D)\geq h(\rho^A)>h(\rho^{AB})$.

It can be easily checked that $C_{(k)}$ in Eq.~\eqref{k-part-entanglement-C0}, $C_{q(k)}$, $C_{\alpha(k)}$, $C_{G,q(k)}$ and $C_{G,\alpha(k)}$ in Eqs.~\eqref{k-part-entanglement-C1}-\eqref{k-part-entanglement-GC2} are not unified. Let 
\beax 
C_{(3)}(|\psi\ra^{ABC}|\psi\ra^{DE}|\psi\ra^{FGH})
=\frac15\left[h(\rho^{AB})+h(\rho^C)+h(\rho^{FG})+h(\rho^H) \right],
\eeax then $C_{(2)}(|\psi\ra^{ABC}|\psi\ra^{DE}|\psi\ra^{FGH})=\frac18(h_A+h_B+h_C+h_D+h_E+h_F+h_G+h_H)$. Hereafter, we denote $h(\rho^X)$ by $h_X$ for simplicity. Clearly, it is not necessary that $C_{(3)}(|\psi\ra^{ABC}|\psi\ra^{DE}|\psi\ra^{FGH})\leq C_{(2)}(|\psi\ra^{ABC}|\psi\ra^{DE}|\psi\ra^{FGH})$. So it is not $k$-monotonic. Let 
\beax 
 C_{(3)}(|\psi\ra^{AB}|\psi\ra^C|\psi\ra^{DEF})=\frac14(h_{AB}+h_C+h_{DE}+h_F)=\frac14(h_{DE}+h_F).
\eeax 
Then $C_{(3)}(|\psi\ra^{AB}|\psi\ra^{DEF})=\frac13(h_{AB}+h_{DE}+h_F)=\frac13(h_{DE}+h_F)$, i.e., $C_{(3)}$ is not coarsening monotonic. In addition $C_{(3)}(|\psi\ra^{AB}|\psi\ra^{C})+C_{(3)}(|\psi\ra^{DEF})=C_{(3)}(|\psi\ra^{DEF})=\frac12(h_{DE}+h_F)$, so $C_{(3)}$ is not additive. For $|\psi\ra^{AB}|\psi\ra^{CD}$, $C_{(2)}(|\psi\ra^{AB}|\psi\ra^{CD})=\frac14(h_A+h_B+h_C+h_D)=\frac12(h_A+h_C)$ may be not larger than $C_{(2)}(AB|C|D)=2h_C/3$.

Take $|\psi\ra^{AB}|\psi\ra^C|\psi\ra^{DE}$, then 
\beax 
&&C_{q(2)}(|\psi\ra^{AB}|\psi\ra^C|\psi\ra^{DE})=\sqrt{\frac25}\sqrt{h_A+h_D}\\
&\neq& C_{q(2)}(|\psi\ra^{AB})+C_{q(2)}(|\psi\ra^C|\psi\ra^{DE})=\sqrt{h_A}+\sqrt{2h_D/3},
\eeax in general. So it is not additive.
For $|\psi\ra^{AB}|\psi\ra^C|\psi\ra^{DEF}$, we assume that 
\beax 
C_{q(3)}(|\psi\ra^{AB}|\psi\ra^C|\psi\ra^{DEF})=\sqrt{h_{DE}+h_F}/2.
\eeax 
But 
\beax 
C_{q(2)}(|\psi\ra^{AB}|\psi\ra^C|\psi\ra^{DEF})=\sqrt{(h_A+h_B+h_D+h_E+h_F)/6},
\eeax which can not guarantee $C_{q(3)}\leq C_{q(2)}$. In addition,
\beax 
 C_{q(3)}(|\psi\ra^{AB}|\psi\ra^C|\psi\ra^{DEF})=\sqrt{h_{F}}/2
<C_{q(3)}(|\psi\ra^{AB}|\psi\ra^{DEF})=\sqrt{2h_{F}/3}
\eeax  
implies that $C_{q(3)}$ is not coarsening monotonic. For $|\psi\ra^{AB}|\psi\ra^{CD}$, $C_{q(2)}(|\psi\ra^{AB}|\psi\ra^{CD})=\sqrt{(h_A+h_C)/2}$ may be not larger than $C_{q(2)}(AB|C|D)=\sqrt{2h_C/3}$, i.e., it is not tightly coarsening monotonic. Replacing $C_{q(k)}$ with $C_{\alpha(k)}$ in arguments above, we get that $C_{\alpha(k)}$ has the same property as that of $C_{q(k)}$.

Consider $|\psi\ra^{AB}|\psi\ra^{CD}$, we have 
\beax 
&&C_{G,q(2)}(|\psi\ra^{AB}|\psi\ra^{CD})=\sqrt{(h_A+h_C)/2}\\
&< &C_{G,q(2)}(|\psi\ra^{AB})+C_{G-q(2)}(|\psi\ra^{CD})=\sqrt{h_A}+\sqrt{h_C}
\eeax  
whenever $h_A h_C>0$. So it is not additive. Let $|\psi_1\ra=|\psi\ra^{ABC}|\psi\ra^D$ with $\rho^A=\rho^B=\rho^C$. It turns out that 
\beax 
C_{G,q(3)}(|\psi_1\ra)=\sqrt[20]{2h_A^{10}/9}>C_{G,q(2)}(|\psi_1\ra)=\sqrt{3h_A/4}.
\eeax 
Namely, it is not $k$-monotonic. Let $|\psi_2\ra=|\psi\ra^{AB}|\psi\ra^C|\psi\ra^D$. Then 
\beax 
C_{G,q(2)}(|\psi_2\ra)=\sqrt{2h_A}/2<C_{G,q(2)}(|\psi\ra^{AB}|\psi\ra^D)=\sqrt{2h_A/3},
\eeax 
i.e., it is not coarsening monotonic. If $h_C>3h_A>0$, $C_{G,q(2)}(|\psi\ra^{AB}|\psi\ra^{CD})=\sqrt{(h_A+h_C)/2}<C_{G,q(2)}(AB|C|D)=\sqrt{2h_C/3}$. So it is not tightly coarsening monotonic either. $C_{G,\alpha(k)}$ has the same property as that of $C_{G,q(k)}$.

By definitions, both $E_{(k)}$ and $E'_{(k)}$ are not genuine $k$-partite entanglement measures.

\section{Examples}\label{sec5}

We illustrate $E_{(k)}$, $\mE_{(k)}$, and $E'_{(k)}$  with the reduced functions $h_C(\rho)=\sqrt{2(1-\tr\rho^2)}$ and  $h(\rho)=S(\rho)$, respectively.
We denote them by $\acute{C}_{(k)}$ (in order to distinguish it from $C_{(k)}$ in Eq.~\eqref{k-part-entanglement-C0}), $\mC_{(k)}$, and $C'_{(k)}$ if the reduced function is $h_C$, and by $E_{(k)}$, $\mE_{(k)}$, and $E'_{(k)}$ whenever the reduced function is $S$. Since $S$ and $h_C$ are subadditive~\cite{Guo2024rip,Wehrl1978}, so  $E_{(k)}$, $\mE_{(k)}$,  $\acute{C}_{(k)}$, and $\mC_{(k)}$ are complete. $E'_{(k)}$ and $C'_{(k)}$ are unified while $E'_{(2)}$ and $C'_{(2)}$ are complete.

Let $|\psi\ra=|GHZ_4\ra^{ABCD}|W_3\ra^{EFG}|\psi\ra^H$ with $|GHZ_4\ra^{ABCD}$ is the four-qubit GHZ state and $|W_3\ra^{EFG}$ is the three-qubit W state.
Then
\beax 
\acute{C}_{(4)}(|\psi\ra)&=&2,\quad
\acute{C}_{(3)}(|\psi\ra)=\acute{C}_{(2)}(|\psi\ra)=2+\sqrt{2},\\
\mC_{(4)}(|\psi\ra)&=&7/2,\quad
\mC_{(3)}(|\psi\ra)=\acute{C}_{(2)}(|\psi\ra)=7/2+\sqrt{2},\\
C'_{(4)}(|\psi\ra)&=&3/2,\quad
C'_{(3)}(|\psi\ra)=1+2\sqrt{2}/3,\quad 
C'_{(2)}(|\psi\ra)=2+\sqrt{2},\\
E_{(4)}(|\psi\ra)&=&2,\quad 
E_{(3)}(|\psi\ra)=E_{(2)}(|\psi\ra)=1+\frac32\log_23,\\
\mE_{(4)}(|\psi\ra)&=&7/2,\quad 
\mE_{(3)}(|\psi\ra)=\mE_{(2)}(|\psi\ra)=5/2+\frac32\log_23,\\
E'_{(4)}(|\psi\ra)&=&3/2,\quad
E'_{(3)}(|\psi\ra)=1/3+\log_23,\quad 
E'_{(2)}(|\psi\ra)=1+\frac32\log_23.
\eeax

For $|\phi\ra=|W_3\ra|\psi^+\ra=\frac{1}{\sqrt{6}}(|100\ra+|010\ra+|001\ra)(|00\ra+|11\ra)$, we have 
\beax 
\acute{C}_{(3)}(|\psi\ra)&=&\sqrt{2},\quad
\acute{C}_{(2)}(|\psi\ra)=1+\sqrt{2},\\
\mC_{(3)}(|\psi\ra)&=&\sqrt{2},\quad
\mC_{(2)}(|\psi\ra)=1+\sqrt{2},\\
C'_{(3)}(|\psi\ra)&=&2\sqrt{2}/3,\quad
C'_{(2)}(|\psi\ra)=1+\sqrt{2},\\
E_{(3)}(|\psi\ra)&=&\frac32\log_23-1,\quad
E_{(2)}(|\psi\ra)=\frac32\log_23,\\
\mE_{(3)}(|\psi\ra)&=&\frac32\log_23-1,\quad
\mE_{(2)}(|\psi\ra)=\frac32\log_23,\\
E'_{(3)}(|\psi\ra)&=&\log_23-\frac23,\quad
E'_{(2)}(|\psi\ra)=\frac32\log_23.
\eeax

\section{Conclusion}\label{sec6}

We have established the axiomatic postulates for complete measure of the $k$-partite entanglement and presented two classes of $k$-partite entanglement measures. Comparing with the axiomatic postulates of complete $k$-entanglement measure in Ref.~\cite{Guo2024pra}, both the unification condition and the hierarchy condition were modified accordingly in order to make them in consistence with the type of entanglement considered. Together with the complete $k$-entanglement measure, we get a further progress in characterizing of multipartite entanglement. In comparison, although the $k$-PEM is far different from the $k$-entanglement measure, it has some similarities to the $k$-entanglement measure: all of them can be defined by the reduced function and in such a sense the completeness is always related to the subadditivity of the reduced function. In addition, we can discuss the monogamy and the complete monogamy relations of the $k$-entanglement measure, but it seems not compatible for $k$-PEM. Going further, our result is applicable for other $k$-partite measure of quantum correlations since it is based on the coarsening relation of partitions.

\begin{acknowledgements}
This work is supported by the National Natural Science Foundation of China under Grant Nos.~12471434 and 11971277, the Program for Young Talents of Science and Technology in Universities of Inner Mongolia Autonomous Region under Grant No. NJYT25010, and the High-Level Talent Research Start-up Fund of Inner Mongolia University under Grant No. 10000-A23207007.
\end{acknowledgements}

\bibliographystyle{spmpsci}      


\end{document}